\documentclass[reviewcopy]{elsart}

\usepackage{amsmath}
\usepackage{amsfonts}
\usepackage{amssymb}
\usepackage [ansinew] {inputenc}
\usepackage{graphicx}
\usepackage[nomarkers,figuresonly]{endfloat}

\begin{document}

\begin{frontmatter}
\title{Digital calculus: \\ Cellular automata dynamics \\
in closed form}
\author{V.~Garc\'{\i}a-Morales\corauthref{cor1}}
\ead{garmovla@uv.es}
\corauth[cor1]{Corresponding author. Tel: +49 15224020629}
\address{Departament de Termodin\`amica, Universitat de Val\`encia, \\ E-46100 Burjassot, Spain}
\begin{abstract}
\noindent{A simple mathematical expression for the universal map for cellular automata is found in closed form with the help of a digit function, whose most basic properties are established. This result is found after proving a theorem on the composition of functions on finite sets. The expression (and the technique used to obtain it) opens the possibility of gaining mathematical insight in any cellular automaton rule since it constitutes at the same time a simple and fast algorithm to implement any such rule.}
\end{abstract}
\end{frontmatter}

\section{Introduction}

Cellular automata (CA) constitute an important tool in the study of complex systems \cite{Wolfram,Adamatzky,Wuensche,Chua1}. CA have been satisfactorily used to model physical \cite{Chopard}, chemical \cite{Kier} and biological processes and, in certain cases, constitute an alternative to continuous models described by partial differential equations \cite{Toffoli}. Natural patterns as those found in some seashells, like the Conus marmorea \cite{Meinhardt} can readily be obtained with CA \cite{Wolfram}. Although such patterns may involve many microscopic degrees of freedom, a coarse graining of them can also be generally described by cellular automata \cite{Israeli1,Israeli2} and the latter can thus be used to capture the collective behavior of spatiotemporal pattern forming systems. 

Despite involving only a finite number $p$ of dynamical states and the interactions having a finite range, already the most elementary CA can generate the highest complexity \cite{Wolfram, VGM3}. The mathematical description of such systems in terms of discrete maps seems to have been only recently attempted \cite{Chua1, Ozelim} and previous researchers have mostly relied upon the computer to get insight in CA behavior. In the monumental work \cite{Chua1}, Chua has pointed out how useful is to have mathematical expressions to gain insight on CA behavior: nonlinear dynamical methods, number theoretical, algebraic and statistical techniques can then be fruitfully applied to these systems. Chua describes a universal neuron which provides a mathematical description for Wolfram's 256 elementary CA \cite{Chua1}. However, this model generally depends on eight adjustable parameters that need to be sought with the computer, and which are then tabulated occupying several pages in \cite{Chua1}.  Ozelim et. al. \cite{Ozelim} have also introduced the so-called iota-delta function to describe Wolfram's elementary CA. Again, however, their model depends on freely adjustable parameters that need to be tabulated separately: in order to describe CA behavior, such numerical parameter values need to be given as input together with the specification of the dynamical state of the cells. In our opinion, the presence of such parameters in the equations is an obvious major drawback from a mathematical point of view.  

\begin{figure}
\includegraphics[width=0.8\textwidth]{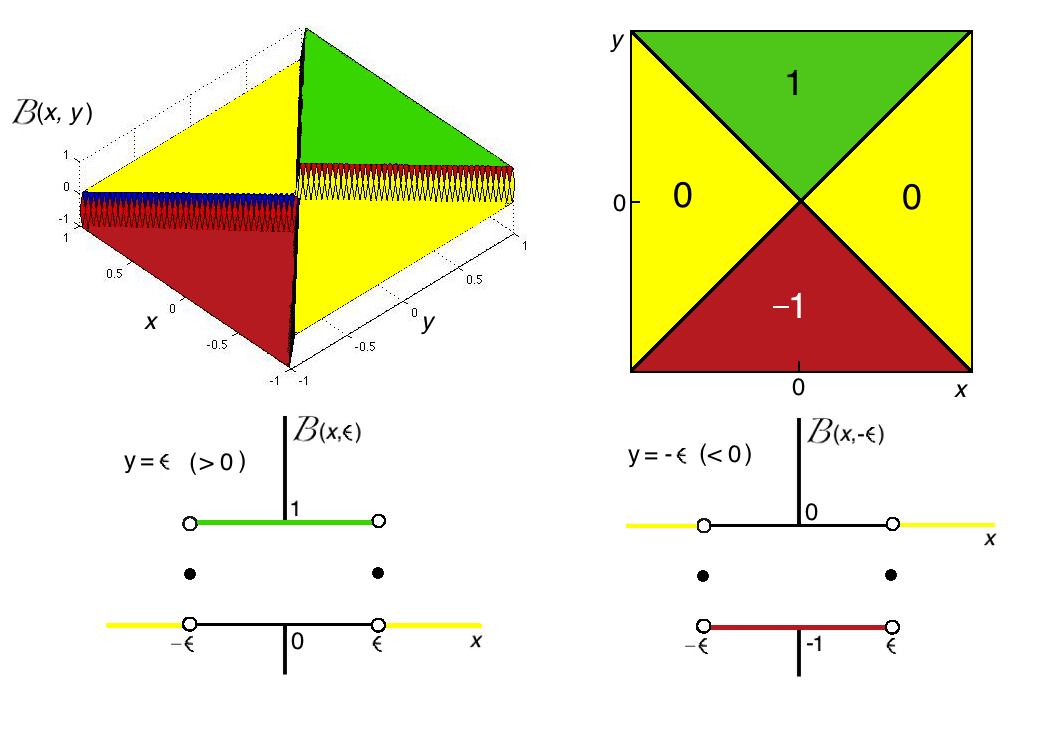}
\caption{The $\mathcal{B}$-function for real-valued arguments (top) and for $y=\pm \epsilon$ constant in the second argument (bottom).} \label{box}
\end{figure}

In \cite{VGM1} we have introduced a universal map for cellular automata which does not depend on any adjustable parameter and which is not only able to simulate Wolfram's 256 elementary CA, but any deterministic CA with any number of dynamical states. This map can also be easily extended to any number of dimensions. In \cite{VGM2} and \cite{VGM3} we have discussed some symmetries of the universal CA map that help to classify the rules into symmetry classes \cite{VGM2} and to understand the kind of symmetry breaking that leads to complexity at this elementary level \cite{VGM3}.  We have termed $\mathcal{B}$-calculus the mathematical techniques introduced in \cite{VGM1} because these strongly rely on a function of two arguments that is called the boxcar or $\mathcal{B}$-function \cite{VGM1}
\begin{equation}
\mathcal{B}(x,y)=\frac{1}{2}\left(\frac{x+y}{|x+y|}-\frac{x-y}{|x-y|}\right) \label{boxy}
\end{equation}
This function constitutes the building block of $\mathcal{B}$-calculus. It returns the sign of the second argument $y$, if $-|y| < x < |y|$, $\text{sign y}/2$ if $x=y$ and '0' otherwise. The $\mathcal{B}$-function is thus an even function of its first argument and an odd function of its second one. The function is represented in Fig. \ref{box}.

As introduced in \cite{VGM1}, the universal map is given in terms of a sum that involves $p^{l+r+1}$ terms, where $p$ is the alphabet size and $l$ and $r$ denote the number of neighbors to the left and to the right (respectively) of the symbol that is updated at the next time step. In this paper we reduce this sum to a single evaluation over the neighborhood values: \emph{the universal map is fully reduced to a closed form with mathematical structure}. The approach makes use of a digit function whose main mathematical properties are first established. Then, we show how the digit function alone is able to describe all deterministic CA in 1D. Extensions to more dimensions also become trivial although we refrain from discussing them here. 


We briefly describe the universal map for cellular automata, whose closed form we then derive. 
Let us consider a 1D ring containing a total number of $N_{s}$ sites. An input is given as initial condition in the form of a vector $\mathbf{x}_{0}=(x_{0}^{1},...,x_{0}^{N_{s}})$. Each of the $x^{j}_{0}$ is an integer in $[0,p-1]$ where superindex $j \in [1, N_{s}]$ specifies the position of the site on the 1D ring. At each $t$ the vector $\mathbf{x}_{t}=(x_{t}^{1},...,x_{t}^{N_{s}})$ specifies the state of the CA.  Periodic boundary conditions are considered so that $x_{t}^{N_{s}+1}=x_{t}^{1}$ and $x_{t}^{0}=x_{t}^{N_{s}}$. Let $x_{t+1}^{j}$ be taken to denote the value of site $j$ at time step $t+1$.  Formally, its dependence on the values at the previous time step is given through the mapping $x_{t+1}^{j}=\ ^{l}R_{p}^{r}(x_{t}^{j+l},\ ... x_{t}^{j}, \ ...,x_{t}^{j-r} )$, which we abbreviate as $x_{t+1}^{j}=\ ^{l}R_{p}^{r}(x_{t}^{j})$ with the understanding that the function on the r.h.s depends on all site values within the neighborhood, with range $\rho=l+r+1$, which contains the site $j$ updated at the next time ($l$ and $r$ denote the number of cells to the left and to the right of site $j$ respectively). We take the convention that $j$ increases to the left. The integer number $n$ in base 10, which runs between $0$ and $p^{r+l+1}-1$, indexes all possible neighborhood values coming from the different configurations of site values. Each of these configurations compares to the dynamical configuration reached by site $j$ and its $r$ and $l$ first-neighbors at time $t$ and given by 
\begin{equation}
n_{t}^{j}=\sum_{k=-r}^{l}p^{k+r}x_{t}^{j+k} \label{NV}
\end{equation}
We will refer to this latter quantity often as the \emph{neighborhood value}. The possible outputs $a_{n}$ for each configuration $n$ are also integers $\in[0,p-1]$. An integer number $R$ can then be given in the decimal base to fully specify the rule $^{l}R_{p}^{r} $ as 
\begin{equation}
R \equiv \sum_{n=0}^{p^{r+l+1}-1}a_{n}p^{n}. \label{RWolf}
\end{equation}
This is the so-called Wolfram code of the CA dynamics. With all these specifications we have the following universal map \cite{VGM1}
\begin{equation}
x_{t+1}^{j}=\ ^{l}R_{p}^{r}(x_{t}^{j})=\sum_{n=0}^{p^{r+l+1}-1}a_{n}\mathcal{B}\left(n-\sum_{k=-r}^{l}p^{k+r}x_{t}^{j+k}, \frac{1}{2} \right)=a_{n_{t}^{j}} \label{CA}
\end{equation}
This map, although describing \emph{all} 1D deterministic cellular automata which are first order in time has the shortcoming that it generally depends on a sum which can contain up to $p^{l+r+1}$ terms. Although Eq. (\ref{CA}) is advantageous in a certain sense because the $\mathcal{B}$-functions in the map behave as an orthonormal base (thus making the transition to a Hilbert space clear, if the $a_{n}$ are allowed to take on complex values) the map is  computationally demanding for $l+r+1$ of $p$ large and a closed form can be desirable in certain cases to gain further mathematical insight and to test the validity of certain calculations with the computer. The finding of such closed form is the goal of the next section.

\section{The digit function}

Let now $p>1$, $p\in \mathbb{N}$. The representation in radix $p$ of $x \in \mathbb{R}$  is a (generally infinite) sequence of the form
\begin{equation}
x=a_{N}p^{N}+a_{N-1}p^{N-1}+\ldots+a_{0}+a_{-1}p^{-1}+a_{-2}p^{-2}+\ldots \label{bexpareal}
\end{equation}
where the $a_{k\in \mathbb{Z}}$'s, the so-called digits of the expansion, are all positive or negative, with absolute values being integers in $S$, the set of integer numbers between $0$ and $p-1$.  Every real number can be represented in this form. Let $\lfloor \ldots \rfloor$ denote the closest lower integer (floor function) and $\lceil \ldots \rceil$ the closest upper integer (ceiling function). The following function
\begin{equation}
\text{sign}(x) p^{-D}\lfloor p^{D}|x|   \rfloor \equiv a_{N}p^{N}+a_{N-1}p^{N-1}+\ldots+a_{0}+a_{-1}p^{-1}+a_{-2}p^{-2}+\ldots +a_{-D}p^{-D} 
\end{equation}
is the \emph{rational truncation to precision $D$ in radix p} of the real number $x$. We thus realize the following interesting fact: the operators $\lfloor \ldots \rfloor$ and multipliying by $p^{D}$ \emph{do not commute} \cite{QUANTUM}, i.e. $\lfloor p^{D}\ldots \rfloor \ne p^{D}\lfloor \ldots \rfloor$. Indeed, we have
\begin{equation}
\text{sign}(x) \left(p^{-D}\lfloor p^{D}|x|   \rfloor- \lfloor |x|   \rfloor\right)=a_{-1}p^{-1}+a_{-2}p^{-2}+\ldots +a_{-D}p^{-D}
\end{equation}
which gives the rational truncation to precision $D$ in radix $p$ of the fractional part of $x$.
Let us consider $x$ nonnegative for simplicity. Now, note that   
\begin{equation}
\left \lfloor \frac{x}{p^{k}} \right \rfloor=a_{N}p^{N-k}+a_{N-1}p^{N-k-1}+\ldots +a_{k+1}p+a_{k} 
\end{equation}
and
\begin{equation}
p\left \lfloor \frac{x}{p^{k+1}} \right \rfloor=a_{N}p^{N-k}+a_{N-1}p^{N-k-1}+\ldots +a_{k+1}p 
\end{equation}
whence, by subtracting both equations, we obtain the digit $a_{k}$.




The digit function, for $p \in \mathbb{N}$, $k \in \mathbb{Z}$ and $x \in \mathbb{R}$ is then defined as \cite{QUANTUM} 
\begin{equation}
\mathbf{d}_{p}(k,x)=\left \lfloor \frac{x}{p^{k}} \right \rfloor-p\left \lfloor \frac{x}{p^{k+1}} \right \rfloor    \label{cucuAreal}
\end{equation}
and gives the $k$-th digit of the real number $x$ (when it is non-negative) in a positional numeral system in radix $p > 1$. If $p=1$ the digit function satisfies $\mathbf{d}_{1}(k,x)=\mathbf{d}_{1}(0,x)=0$ and it does not relate to a positional numeral system.

With the digit function we can express Eq. (\ref{bexpareal}) as 
\begin{equation}
x=\text{sign}(x)\sum_{k=-\infty}^{\lfloor \log_{p}|x| \rfloor} p^{k} \mathbf{d}_{p}(k,|x|) \label{idenreal}
\end{equation}
Note that the digit function is also defined for $x$ negative since in that case we have, from the definition 
\begin{equation}
\mathbf{d}_{p}(k,-x)=p\left \lceil \frac{x}{p^{k+1}}\right \rceil-\left \lceil \frac{x}{p^{k}} \right \rceil =
\mathbf{d}_{p}(k,p^{k+1}-x)
\end{equation}
where we have used that $\lfloor -y \rfloor=-\lceil y \rceil$ (for any real number $y$).

\begin{figure*} 
\includegraphics[width=0.7\textwidth]{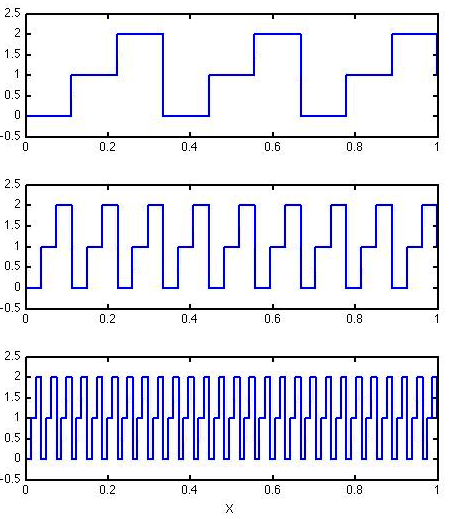}
\caption{The digit function $\mathbf{d}_{3}(k,x)$ plotted for $k=-2, -3, -4$ (from top to bottom) in the interval $x \in [0,1]$} \label{digitfu}
\end{figure*}

Most of the properties of the digit function are made apparent by plotting $\mathbf{d}_{p}(k,x)$ for a fixed value of $p$ as a function of $x$ and different values for $k$, as shown in Fig. \ref{digitfu}, where $\mathbf{d}_{3}(k,x)$ is shown for $k=-2, -3, -4$ in the interval $x \in [0,1]$. The digit function is a staircase of $p$ levels taking discrete integer values between $0$ and $p-1$. Each time that $x$ is divisible by $p^{k+1}$ the ascent of the staircase is broken and the level is  set again to zero and a new staircase begins. The next proposition formalizes some of these properties.

\noindent \textbf{Proposition 1.} \emph{The digit function satisfies (for $n$ and $m$ nonnegative integers)}
\begin{eqnarray}
\mathbf{d}_{p}(k,x+np^{k+1}) &=& \mathbf{d}_{p}(k,x) \label{pro1}\\
\mathbf{d}_{p}(k,p^{k}x) &=&\mathbf{d}_{p}(0,x) \label{pro3} \\
\mathbf{d}_{p}(0,\mathbf{d}_{p}(k,x)) &=& \mathbf{d}_{p}(k,x) \label{pro2} \\
\mathbf{d}_{p}(0,n+\mathbf{d}_{p}(0,m)) &=& \mathbf{d}_{p}(0,n+m) \label{pro2b} \\
\mathbf{d}_{p}(0,n\mathbf{d}_{p}(0,m)) &=& \mathbf{d}_{p}(0,nm) \label{pro2c} \\
 \mathbf{d}_p(n,p^{m}) &=& \mathbf{d}_p(m,p^{n})= \mathcal{B}\left(m-n, \frac{1}{2} \right)  \label{krone} \\
 \mathbf{d}_p(k,n) &=& \sum_{j=0}^{m>n}\mathbf{d}_p(k,j)\mathbf{d}_p(j,p^{n}) \label{expan}
\end{eqnarray}

 
\noindent \emph{Proof} Eq. (\ref{pro1}) follows directly from the definition. We have
\begin{eqnarray}
\mathbf{d}_{p}(k,x+np^{k+1})&=&\left \lfloor \frac{x+np^{k+1}}{p^{k}} \right \rfloor-p\left \lfloor \frac{x+np^{k+1}}{p^{k+1}} \right \rfloor =\left \lfloor \frac{x}{p^{k}}+np \right \rfloor-p\left \lfloor \frac{x}{p^{k+1}}+n \right \rfloor \nonumber \\
&=&\left \lfloor \frac{x}{p^{k}} \right \rfloor+np-p\left \lfloor \frac{x}{p^{k+1}} \right \rfloor-np=
\left \lfloor \frac{x}{p^{k}} \right \rfloor-p\left \lfloor \frac{x}{p^{k+1}} \right \rfloor \equiv \mathbf{d}_{p}(k,x)
\end{eqnarray}
Eq. (\ref{pro3}) is also trivial to prove
\begin{eqnarray}
\mathbf{d}_{p}(k,p^{k}x)&=&\left \lfloor \frac{p^{k}x}{p^{k}} \right \rfloor-p\left \lfloor \frac{p^{k}x}{p^{k+1}} \right \rfloor =\left \lfloor x \right \rfloor-p\left \lfloor \frac{x}{p} \right \rfloor=\mathbf{d}_{p}(0,x) \nonumber
\end{eqnarray} 

Eq. (\ref{pro2}) follows by noting that $\mathbf{d}_{p}(k,x)$ is an integer $0 \le \mathbf{d}_{p}(k,x) \le p-1$. Then 
\begin{equation}
\mathbf{d}_{p}(0,\mathbf{d}_{p}(k,x))=\mathbf{d}_{p}(k,x)-p\left \lfloor \frac{\mathbf{d}_{p}(k,x)}{p} \right \rfloor=\mathbf{d}_{p}(k,x)
\end{equation}
since $\left \lfloor \mathbf{d}_{p}(k,x)/p \right \rfloor=0$. Eq. (\ref{pro2b}) is proved by using Euclidean division, since we can always write $m=ap+b$ with $a, b$ integers and $b=\mathbf{d}_{p}(0,m)$. Therefore
\begin{equation}
\mathbf{d}_{p}(0,n+\mathbf{d}_{p}(0,m))=\mathbf{d}_{p}(0,n+ap+\mathbf{d}_{p}(0,m))=\mathbf{d}_{p}(0,n+m)  
\end{equation}
where Eq. (\ref{pro1}) has also been used. Eq. (\ref{pro2c}) is also proved in a similar way, using the distributive property of ordinary addition and multiplication as well. Eq. (\ref{krone}) is simply proved from the definition and Eq. (\ref{expan}) is a simple consequence of it. $\Box$

\begin{figure*} 
\includegraphics[width=0.8 \textwidth]{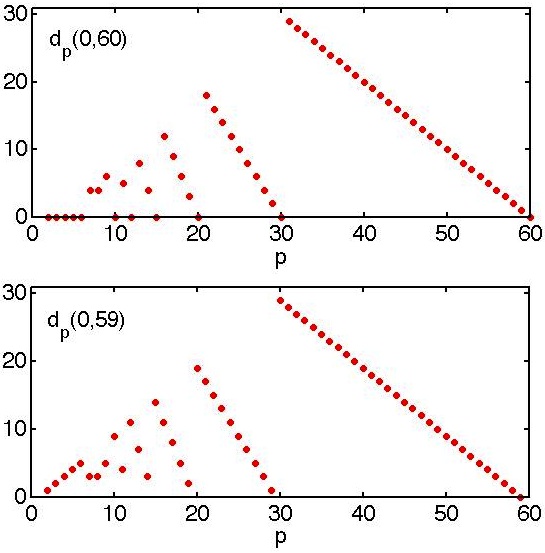}
\caption{The digit function $\mathbf{d}_{p}(0,60)$ (top) and $\mathbf{d}_{p}(0,59)$ (bottom) plotted vs. $p$.} \label{digitvsp}
\end{figure*}

The zeroth digit of the non-negative integer $n$ in any radix $p$, $\mathbf{d}_{p}(0,n)$, governs the divisibility of $n$ by the radix and yields the remainder of that division. In Fig. (\ref{digitvsp}) $\mathbf{d}_{p}(0,60)$ (top) and $\mathbf{d}_{p}(0,59)$ (bottom) are plotted vs. $p$. There are several interesting features that we notice immediately. In the case $n=60$ the function touches the abscissa at the radix values that divide $60$. Such intersections are absent in the case $n=59$ (save, of course, $p=59$) because $59$ is prime. Of course, $\prod_{p=2}^{A-1}\mathbf{d}_{p}(0,A) \ne 0$ only if $A$ is prime.

We also see linearly decreasing branches with different slopes after a certain threshold value of $p$. This behavior is readily understood by the definition of the digit function. Note that, from Eq. (\ref{cucuAreal}), when $p$ is sufficiently large the digit function reduces to
\begin{eqnarray}
\mathbf{d}_{p}(0,n)&=&n-p \ \qquad \text{if $n/2< p \le n$} \nonumber \\ 
&=&n-2p \qquad \text{if $n/3 <    p \le n/2$} \nonumber \\
&=&n-3p \qquad \text{if $n/4 <    p \le n/3$} \nonumber 
\end{eqnarray}
etc. The function values are all contained within a triangle formed by the lines $f_{1}(p)=p-1$ and $f_{2}(p)=n-p$. The following property follows easily
\begin{equation}
\prod_{p=2}^{n}\mathbf{d}_{p}(0,p-1)=\prod_{p=2}^{n}(p-1)=(n-1)! \label{Willsein}
\end{equation}

\section{Main results}

More significant digits exhibit even a more complex and non-monotonic behavior. Remarkably, though, those significant digits can be used to split the composition of two functions over finite sets. This is the result contained in the following theorem.

\noindent \textbf{Theorem 1.} \textbf{(Composition-decomposition theorem.)} Let  $g: S^{N} \to A$ and $f:A \to S$ be two functions on finite sets $S$ consisting of all integers $\in [0,p-1]$ and $A$ consisting of all integers $\in [0,p^{N}-1]$ with $p,\ N \in \mathbb{N}$ and let $k \in S^{N}$. We have
\begin{equation}
f(g(k))=\mathbf{d}_{p}\left(g(k), \sum_{n=0}^{p^{N}-1}p^{n}f(n)\right)=\left \lfloor \frac{\sum_{n=0}^{p^{N}-1}p^{n}f(n)}{p^{g(k)}} \right \rfloor-p \left \lfloor \frac{\sum_{n=0}^{p^{N}-1}p^{n}f(n)}{p^{g(k)+1}} \right \rfloor \label{condecon}
\end{equation}

\noindent \emph{Proof.} From Eq. (\ref{pro3}) we have
\begin{equation}
\mathbf{d}_{p}\left(g(k), \sum_{n=0}^{p^{N}-1}p^{n}f(n)\right)=\mathbf{d}_{p}\left(0, \sum_{n=0}^{p^{N}-1}p^{n-g(k)}f(n)\right)
\end{equation}
This function extracts the zeroth digit of the quantity $\sum_{n=0}^{p^{N}-1}p^{n-g(k)}f(n)$, i.e. the value of $f(n)$ for which the exponent of the accompanying power of $p$ within the sum is zero. But this happens when $n=g(k)$ from which the result follows. $\Box$

We have now all what we need to prove our main result.

\noindent \textbf{Theorem 2.} \textbf{(Universal map for CA in closed form.)} \emph{Eq. (\ref{CA}) is equivalent to}
\begin{equation}
x_{t+1}^{j}=\mathbf{d}_{p}\left(\sum_{k=-r}^{l}p^{k+r}x_{t}^{j+k} , R \right)=\left \lfloor \frac{R}{p^{\sum_{k=-r}^{l}p^{k+r}x_{t}^{j+k}}} \right \rfloor-p \left \lfloor \frac{R}{p^{1+\sum_{k=-r}^{l}p^{k+r}x_{t}^{j+k}}} \right \rfloor 
\label{themap}
\end{equation}

\noindent \emph{Proof.} Let us take $N=l+r+1$. Eq. (\ref{CA}) can be seen as the composition of two functions. The first one $g(k)=\sum_{k=-l}^{r}p^{k+r}x_{t}^{j+k}$ sends the $N$ integers $x_{t}^{j+k}$ (each $\in [0,p-1]$) to one integer $n_{t}^{j}\in [0,p^{N}-1]\equiv [0,p^{l+r+1}-1]$ and hence qualifies as a valid function $g(k)$ for the composition-decomposition theorem. The second one, $f(n)=a_{n}$ sends the integers $n \in [0,p^{N}-1]$ to one integer $\in [0,p-1]$. Thus, by the composition decomposition theorem we have, by replacing $g(k)$ and $f(n)$ under this identification
\begin{equation}
x_{t+1}^{j}=\mathbf{d}_{p}\left(\sum_{k=-r}^{l}p^{k+r}x_{t}^{j+k} , \sum_{n=0}^{p^{l+r+1}}p^{n}a_{n} \right)
\end{equation}
from which, by using Eq. (\ref{RWolf}), the result follows. $\Box$

The above result replaces a sum over $p^{l+r+1}-1$ terms by \emph{just only one term} evaluated on the neighborhood of the site at location $j$. Furthermore, \emph{we do not need to give any table of configurations as input but just only the Wolfram code (a nonnegative integer $R \in [0, p^{p^{l+r+1}}])$ and the neighborhood parameters $p$, $l$ and $r$ directly}. For example, the celebrated Wolfram's rule 110 takes the simple form
\begin{equation}
x_{t+1}^{j}=\mathbf{d}_{2}\left(\sum_{k=-1}^{1}2^{k+r}x_{t}^{j+k} , 110 \right)
\end{equation}
The maps for all Wolfram's 256 CA with $l=r=1$ and $p=2$ are thus simply obtained as
\begin{equation}
x_{t+1}^{j}=\mathbf{d}_{2}\left(\sum_{k=-1}^{1}2^{k+r}x_{t}^{j+k} , R \right)
\end{equation}
by giving an input $R \in [0,255]$. 

\begin{figure*} 
\includegraphics[width=1.0 \textwidth]{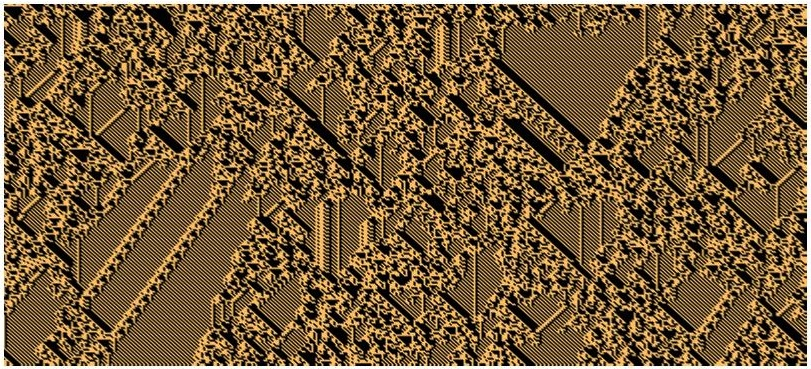}
\caption{Spatiotemporal evolution of the rule $x_{t+1}^{j}=\mathbf{d}_{2}\left(\sum_{k=-2}^{2}2^{k+r}x_{t}^{j+k} , 1771466136 \right)$ starting from an arbitrary initial condition on a ring of 400 sites and 200 time steps. Time flows from top to bottom.} \label{rula}
\end{figure*}

As a further illustration, the rule $x_{t+1}^{j}=\mathbf{d}_{2}\left(\sum_{k=-2}^{2}2^{k+r}x_{t}^{j+k} , 1771466136 \right)$ is plotted in Fig. \ref{rula} starting from an arbitrary initial condition. This rule displays many coherent soliton-like structures that are created and annihilated against a quiescent background.

With Eq. (\ref{themap}) alone we are able to explore the whole CA rule space within computational reach and go even beyond, because \emph{the mathematical expression is valid for arbitrarily large parameter values} regardless of our computational power to carry out the simulations of a given rule.

\bibliography{biblos}{}
\bibliographystyle{h-physrev3.bst}

\end{document}